\documentclass[runningheads,a4paper]{llncs}

\usepackage{amsmath}
\usepackage{amssymb}
\usepackage{graphicx}

\usepackage{url}
\urldef{\mailsa}\path|hed_up@iasystems.org|
\urldef{\mailsb}\path|lsb@dia.uned.es|

\newcommand{\keywords}[1]{\par\addvspace\baselineskip
\noindent\keywordname\enspace\ignorespaces#1}

\begin{document}

\mainmatter  

\title{Bayesian Unbiasing of the {\it Gaia} space mission time series database}
\titlerunning{Bayesian unbiasing of the {\it Gaia} time series catalog}
\author{H\'{e}ctor E. Delgado%
\and Luis M. Sarro}
\institute{Dpto. de Inteligencia Artificial , UNED, Juan del Rosal,
	16, 28040 Madrid, Spain\\ 
	\mailsa\\
	\mailsb\\
    }

\maketitle

\begin{abstract}
21$^{st}$ century astrophysicists are confronted with the
  herculean task of distilling the maximum scientific return from
  extremely expensive and complex space- or ground-based instrumental
  projects. This paper concentrates in the mining of the time series
  catalog produced by the European Space Agency {\sl Gaia} mission,
  launched in December 2013. We tackle in particular the problem of
  inferring the true distribution of the variability properties of
  Cepheid stars in the Milky Way satellite galaxy known as the Large
  Magellanic Cloud (LMC). Classical Cepheid stars are the first step
  in the so-called distance ladder: a series of techniques to measure
  cosmological distances and decipher the structure and evolution of
  our Universe.  In this work we attempt to unbias the catalog by
  modelling the aliasing phenomenon that distorts the true
  distribution of periods.  We have represented the problem by a
  2-level generative Bayesian graphical model and used a Markov chain
  Monte Carlo (MCMC) algorithm for inference (classification and
  regression). Our results with synthetic data show that the system
  successfully removes systematic biases and is able to infer the true
  hyperparameters of the frequency and magnitude distributions.
  

\end{abstract}

\section{Introduction}

{\sl Gaia} \cite{Gaia2016} is a European Space Agency (ESA) space
  mission, launched in December 2013, whose main objective is to compile
  a large-scale astronomical survey of about one billion stars
  ($\approx$1\%) of our Galaxy and its Local Group. The satellite will
  scan the entire sky for about 5 years yielding an unprecedented
  catalog in both size and precision of positions, distances and
  proper motion measures. Additionally, it will perform multi-epoch
  photometry (70 transits per object on average) which renders the
  satellite suitable too for studies of stellar variability.
  Amongst the many variability types present in the stellar zoo,
  one in particular is of paramount importance: the Classical
  Cepheids. Classical Cepheids represent the first calibrator in the
  cosmic distance ladder used to infer the structure and evolution of
  our Universe, and our current knowledge about the Big Bang, the
  inflationary period, the dark matter problem or dark energy relies
  on the period-luminosity relation for Classical Cepheids
  \cite{Sandage2004}. Therefore, a precise and accurate understanding
  of the population of Classical Cepheids is central to all
  cosmological studies.

  In this paper we address the problem of inferring the true
  properties of this population of variable stars from the
  petabyte-size {\it Gaia} catalog. In order to populate the catalog,
  and as part of a much larger framework to deliver a data set of
  scientific quality, the Data Processing and Analysis Consortium
  (DPAC\footnote{The DPAC (Data Processing and Analysis Consortium) is
    the consortium responsible for building and making accessible the
    GAIA catalogue.}) developed a pipeline to characterize the time
  series observed and classify them. A key element of this process is
  that the time sampling of stellar brightness time series will have
  the imprint of the satellite intrinsic frequencies (amongst other,
  the spinning and precessing frequencies, a description of which is
  out of the scope of this paper). As a consequence, some (but not
  all) of the derived frequencies will be affected by aliasing which
  results in biased samples.

The objective is to characterize the phenomenon of
\emph{aliasing} in the {\sl Gaia} catalog, correct for it, and
reconstruct the real distribution of LMC Classical Cepheids
properties. In order to achieve these goals, we tackle the problem
under the Bayesian paradigm \cite{Gelman2008,Gelman2013} and adopt
the knowledge representation language of Bayesian Networks (BN)
\cite{Pearl1988,Lauritzen1996}. This framework allows a hierarchical
representation of the problem in which the time series gathered by
Gaia are the product of a generative process which ultimately depends
on the parameters of the population of stars. Given that the
computation of the posterior probabilities of our model are
analytically intractable, the inference mechanism of our
proposal is founded in Markov chain Monte Carlo (MCMC) simulation
techniques \cite{Robert2004}.

We have validated our models using a data base of 36688 synthetic
Classical LMC Cepheids time series generated according to controlled
prescriptions based on current understanding of the true
distributions and the satellite characteristics. Our results prove
that we are ready for the second {\sl Gaia} data release expected
for 2018. This will be the first data release to include photometric
  time series (although this still needs to be confirmed).
  
The structure of the rest of the paper is as follows. In section
\ref{sec:BGM-and-Computation} we describe our model and the MCMC
technique used for the inference of the parameters of interest. In
Section \ref{sec:BGM-validation} we validate the model with the
simulated data base in a scenario of extreme aliasing and
  describe the results of this validation procedure.
Finally, in Section \ref{sec:conclusions} we summarize the
contributions of this work and some of its limitations, and give
pointers to future developments.

\section{Hierarchical Modelling of the distribution of pulsation properties of Classical Cepheid Variable stars}\label{sec:BGM-and-Computation}

\subsection{The Hierarchical Model}\label{sec:BGM} 

\begin{figure}[ht]
	\centering
        \includegraphics[scale=0.50]{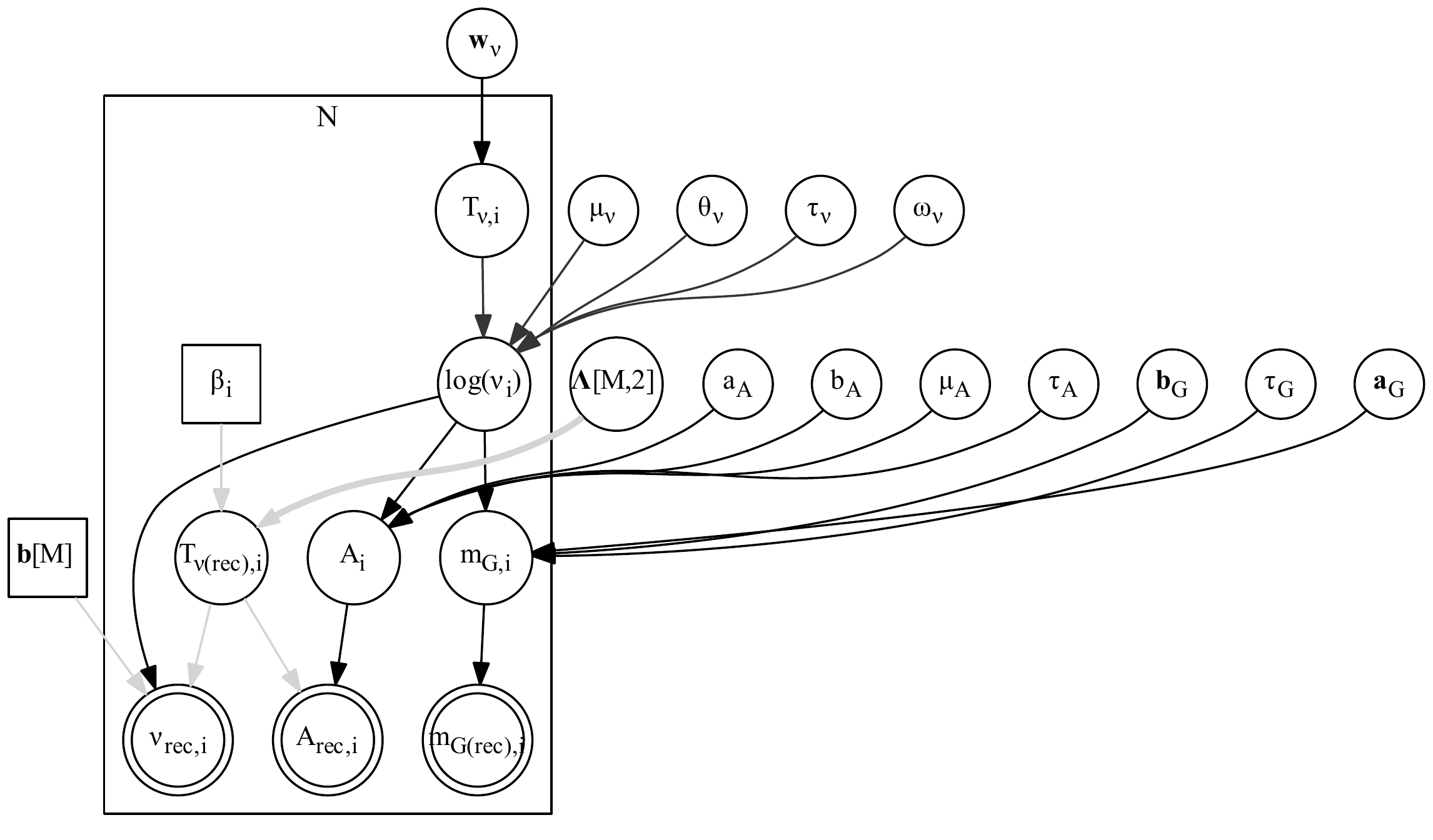}

	\caption{Graph structure of our proposed Bayesian
            Graphical Model (BGM). Most fixed parameters are not
          included in the graph, with the exception of those enclosed
          inside a square. See the text and Table
          \ref{tab:Description-of-parameters.} for node
          descriptions. }
	\label{fig:graph-structure}	
\end{figure}

Figure \ref{fig:graph-structure} and Table
\ref{tab:Description-of-parameters.} depict the structure of the DAG
associated to the model and summarize the meanings of the nodes and
the types of their distributions. We classify the nodes into a
hierarchy of three levels. The hierarchy distinguishes between
evidential nodes (observations), the rest of nodes inside the
rectangle or \emph{plate}, which is replicated \emph{N} times (one
per star), and the nodes outside the rectangle. In the following
paragraphs we describe the parameters and probability distributions
for each level and its contribution to the joint probability
distribution.

\begin{table}
	\caption{Description of parameters. NI = non
          informative\label{tab:Description-of-parameters.}}
        \centering
		\begin{tabular}{ccc}
			\hline 
			\textbf{Node} & \textbf{Description} & \textbf{Type of distribution} \\ 
			\hline 
			$\tau_{G}$ & Precision & Gamma NI prior \\ 
			$\mathbf{a}_{G}$ & Slopes & Gaussian NI prior \\ 
			$\boldsymbol{b}_{G}$ & Intercepts & Gaussian NI prior \\ 
			$m_{G,i}$ & Input apparent G magnitude  & Gaussian \\ 
			$m_{G,\mathrm{rec},i}$ & Recovered apparent G magnitude  & Gaussian \\ 
			\hline 
			$\mu_{A}$ & Mean & Gaussian NI prior \\ 
			$\tau_{A}$ & Precision & Gamma NI prior \\ 
			$a_{A}$ & Slope & Gaussian NI prior \\ 
			$b_{A}$ & Intercept & Gaussian NI prior \\ 
			$A_{i}$ & Input amplitude & Gaussian \\ 
			$A_{\mathrm{rec},i}$ & Recovered amplitude & Mixture of skewed Cauchy \\ 
			\hline 
			$\boldsymbol{w}_{\nu}$ & Mixing proportions. & Gamma NI prior \\ 
			$T_{\nu_{i}}$ & Category of $\log\left(\nu_{i}\right)$ & Categorical \\ 
			$\mu_{\nu}$ & Mean & Non informative \\ 
			$\boldsymbol{\theta}_{\nu}$ & Mean Perturbations & Gaussian NI prior \\ 
			$\tau_{\nu}$ & Precision & Non informative \\ 
			$\boldsymbol{\omega}_{\nu}$ & Precision Perturbations & Uniform prior \\ 
			$\log\left(\nu_{i}\right)$  & Input frequency $\left[d^{-1}\right]$. & Mixture of Gaussian \\ 
			$\Lambda$ & Logistic R. coefficients & Student t prior \\ 
			$T_{\nu_{\mathrm{rec},i}}$ & Category of $\nu_{\mathrm{rec},i}$ & Categorical \\ 
			$\nu_{\mathrm{rec},i}$  & Recovered frequency & Mixture of Gaussian \\ 
			\hline 
		\end{tabular}
\end{table}

\subsubsection{Likelihood.}
In the bottom level of our graph we present the \emph{evidential
  nodes}, that is, the variables measured directly or derived by
  the DPAC
\begin{equation}\label{eq:model-Data}
\mathcal{D}=\left(\nu_{\mathrm{rec},i},A_{\mathrm{rec},i},m_{G_{\mathrm{rec}},i}\right) \;  . 
\end{equation}
These nodes, depicted by double circles, are the output/recovered
frequency $\nu_{\mathrm{rec},i}$, the amplitude $A_{\mathrm{rec},i}$
and the apparent G-magnitude $m_{G_{\mathrm{rec}},i}$ for the
\emph{i}-th star.

\paragraph{Recovered Frequencies.}
Most of the pairs
$\left(\nu_{\mathrm{input}},\nu_{\mathrm{rec}}\right)$ in the
simulated data base fall on straights lines of the form:
\begin{equation}\label{eq:locus-of-recovered-frequencies-general-equation}
\nu_{\mathrm{rec}}=\pm\nu_{\mathrm{input}}\pm k_{1}\nu_{s}\pm
k_{2}\nu_{p} \;  , 
\end{equation}
where $k_{1}\in\left\{ 0,3,7\right\} $, $k_{2}\in\left\{
0,...,19\right\} $, $\nu_{s}\approx\frac{1}{0.25}=4\mathrm{d}^{-1}$ is
the rotational frequency of Gaia and
$\nu_{p}=\frac{1}{63}\mathrm{d}^{-1}$ is its precessional
frequency. We refer to each line as a \emph{locus}/\emph{category} of
recovered frequencies. Excluding the line
$\nu_{\mathrm{rec}}=\nu_{\mathrm{input}}$, all these \emph{loci} 
  correspond to spurious (aliased) frequencies. Based on that, we
parameterize the $i$-th recovered frequency as the following mixture
of Gaussian distributions
\begin{equation}\label{eq:rec-frequency-density}
	f\left(\nu_{\mathrm{rec},i}\mid\log\left(\nu_{i}\right),T_{\nu_{\mathrm{rec},i}}\right)=
	\sum_{j=1}^{M}\delta_{T_{\nu_{\mathrm{rec},i}}}^{j}\mathsf{N}\left(\left(-1\right)^{j-1}10^{\log\left(\nu_{i}\right)}+b_{j},\tau_{\nu_{\mathrm{rec}}}\right) \;  .
\end{equation}
In Equation \ref{eq:rec-frequency-density} the Kronecker deltas
$\delta_{T_{\nu_{\mathrm{rec},i}}}^{j}$ dictate the Gaussian component
to which $\nu_{\mathrm{rec},i}$ belongs according to the value of the
categorical variable $T_{\nu_{\mathrm{rec},i}}$ (described in Section
\ref{subsub:1st-level-of-parameters}). The mean of each component
represents the \emph{locus} in which the input frequency has been
recovered, i.e. the identity locus, with $b_{j}=0$ for $j=1$, or some
locus of spurious (aliased) frequencies for $j>1$. We assume the same
precision $\tau_{\nu_{\mathrm{rec}}}=10000$ for all components.

\paragraph{Recovered Amplitudes.}

To gain insight into the form of the conditional distribution of
the recovered amplitude given the input amplitude we have checked the
hypothesis that recovered amplitudes are also biased by the
  aliasing phenomenon, just as recovered frequencies are. By
analysing the relationship between \emph{loci} of frequencies and pairs
$\left(A_{\mathrm{input}},A_{\mathrm{rec}}\right)$, we have discovered
that for a perfect recovery the distribution $A_{\mathrm{rec}}\mid A$
is skewed to lower amplitudes with a central parameter approximately
equal to the input amplitude. Otherwise, for \emph{loci} of aliased
frequencies we have observed that the skewness of the recovered
amplitude increases as the input amplitude does according to a certain
slope to be determined as part of the model. To account for
  this fact we have fitted two linear regression models
\begin{equation}\label{eq:rec-amplitude-regressions}
A_{\mathrm{rec},i}=\beta_{1}^{j}A_{\mathrm{in},i}+\beta_{0}^{j}+\epsilon_{i}^{j},\;
j=1,2 \; ,
\end{equation}
with $j=1$ corresponding to the identity locus and $j=2$ to
the \emph{loci}
$\nu_{\mathrm{rec}}=\pm\nu_{\mathrm{in}}+7\nu_{s}-3\nu_{p}$
\footnote{We only select these particular \emph{loci} of aliased frequencies because they are the most frequent \emph{loci}  located far away from the identity \emph{locus} and because the model does not work well if we include more \emph{loci} located close to them.}. For the identity locus,
we have assumed a skewed Student t distribution
\cite{Azzalini2008} with one degree of freedom (skewed
Cauchy) for the error component
$\epsilon_{i}^{1}\sim\mathsf{st}\left(0,\omega,\alpha,1\right)$
where $\omega$ and $\alpha$ denote respectively the shape and
scale parameters. For the \emph{locus}
$\nu_{\mathrm{rec}}=\pm\nu_{\mathrm{in}}+7\nu_{s}-3\nu_{p}$
we have assumed that
$\epsilon_{i}^{2}\sim\mathsf{t}\left(0,\omega,1\right)$. Based
on that, we model the conditional distribution for the
recovered amplitude $A_{\mathrm{rec},i}$ by means of the
mixture of two skewed Student t distributions
\begin{equation}\label{eq:rec-amplitude-density}
	\begin{split}
		f\left(A_{\mathrm{rec},i}\mid A_{i},T_{\nu_{\mathrm{rec},i}}\right) & =
		\delta_{T_{\nu_{\mathrm{rec},i}}}^{1}\mathsf{ST}\left(A_{i},0.020,-2.395,1\right)\\
		& +\sum_{j=2}^{M}\delta_{T_{\nu_{\mathrm{rec},i}}}^{j}\mathsf{ST}\left(0.749\cdot A_{i},0.0266,0,1\right) \;  ,
	\end{split}
\end{equation}
where the location parameters $\xi_{1}=A_{i}$, $\xi_{j}=0.749\cdot A_{i},\forall j=2,..,M$, the scale $\omega$ and the shapes $\alpha$ have been obtained from the fitting of the two linear models of Equation
\ref{eq:rec-amplitude-regressions} and taken as constants in our BGM.

\paragraph{Recovered Apparent Magnitudes.}

We parameterize the distribution of the \emph{i}-th recovered apparent
G magnitude by means of a Gaussian distribution with mean $m_{G,i}$
and precision $\tau_{G(rec)}=2.5\mathrm{E}{+5}$ (to be adjusted
  when real {\sl Gaia} data become available)
\begin{equation}\label{eq:rec-ap-magnitude-density}
	f\left(m_{G_{rec},i}\mid
        m_{G,i}\right)=\mathsf{N}\left(m_{G,i},\tau_{G_{rec}}\right)
        \; .
\end{equation}

The conditional distribution of the data given their parents is then given by
\begin{equation}\label{eq:joint-PDF-likelihood-factor}
	\begin{split}
		p\left(\mathcal{D}\mid\boldsymbol{\theta}_{1}\right)  = & \prod_{i=1}^{N}f_{1}\left(\nu_{\mathrm{rec},i}\mid\log\left(\nu_{i}\right),T_{\nu_{\mathrm{rec},i}}\right)\cdot
		f_{2}\left(A_{\mathrm{rec},i}\mid A_{i},T_{\nu_{\mathrm{rec},i}}\right)\\ & \cdot f_{3}\left(m_{G_{\mathrm{rec}},i}\mid m_{G,i}\right) \;  .
	\end{split} 
\end{equation}

\subsubsection{First Level Random Parameters.}\label{subsub:1st-level-of-parameters}

These are
	\begin{equation}\label{eq:model-1s-level-parameters}
	\boldsymbol{\theta}_{1}=\left(\log\left(\nu_{i}\right),A_{i},m_{G,i},T_{\nu_{\mathrm{rec},i}},T_{\nu_{i}}\right) \;  .
	\end{equation}
	
In $\boldsymbol{\theta}_{1}$, we distinguish two classes of nodes. The
\emph{input nodes} are, for the \emph{i}-th star, the real frequency
$\log\left(\nu_{i}\right)$, the real amplitude $A_{i}$ and the real
apparent G-magnitude $m_{G,i}$. The \emph{categorical nodes}
$T_{\nu_{\mathrm{rec},i}}$ and $T_{\nu_{i}}$ determine the component
of a node modelled by a mixture of distributions. $T_{\nu_{i}}$ and
$T_{\nu_{\mathrm{rec},i}}$ are respectively associated with the real
frequency and the recovered frequency and amplitude. In Figure
\ref{fig:graph-structure} all the nodes at this level replicate with
the plate. They depend on (amongst other) non informative orphan
nodes outside the plate.

\paragraph{Categories of Recovered Frequencies.}

The node $T_{\nu_{\mathrm{rec},i}}$ takes a value $j\in\left\{
1,..,M\right\} $ if the \emph{i}-th frequency has been recovered in
the \emph{j}-th locus, which occurs with a probability
$\pi_{ij}$. In this paper we assume that the main factor
  determining the aliasing phenomenon in Gaia is the ecliptic
latitude $\beta$ of the stars. The influence of $\beta$ over the rate
of correct detections of periodic signals by {\it Gaia} has been
studied in \cite{Eyer2005} where it is shown that for high values of
$\beta$, typical of LMC sources, the relation between the rate
of correct detections and $\beta$ is approximately linear with a
negative slope. Based on that, we make $\pi_{ij}$ depend on the
ecliptic latitude $\beta_{i}$ and parameterize this dependence by a
multinomial logistic regression submodel with a \emph{softmax}
transfer function. We model the conditional distribution of
$T_{\nu_{\mathrm{rec},i}}$ as
\begin{equation}\label{eq:category-of-rec-freq-distribution}
p\left(T_{\nu_{\mathrm{rec},i}}\mid\left\{ \boldsymbol{\lambda}_{j}\right\} _{j=2}^{M}\right)=
\mathrm{\mathsf{Cat}}\left(M,\left\{ \pi_{ij}\left(\beta_{i}',\boldsymbol{\lambda}_{j}\right)\right\} _{j=1}^{M}\right) \;  ,
\end{equation}
with
\begin{equation}\label{eq:softmax-recovery-probabilities}
\pi_{ij}\left(\beta_{i}',\boldsymbol{\lambda}_{j}\right)=\frac{e^{\boldsymbol{\lambda}_{j}^{T}\cdot\left(1,\beta_{i}'\right)}}{\sum_{l=1}^{M}e^{\boldsymbol{\lambda}_{l}^{T}\cdot\left(1,\beta_{i}'\right)}} \;  ,
\end{equation}
where we have rescaled the predictor $\beta_{i}$ by subtracting the
mean and dividing by two times the standard deviation,
i.e. $\beta_{i}'=\frac{\beta_{i}-\overline{\beta}}{2\cdot\mathrm{sd}\left(\beta\right)}$,
which guaranties that the mean and the standard deviation are
respectively 0 and 0.5.

\paragraph{Input Frequencies and Categories.}

The marginal distribution of the (decadic) logarithm of the
input frequency in the synthetic data set created by the DPAC Quality
Assessment group was sampled from a mixture of five Gaussian
distributions \cite{Antonello2002}. In our BGM, we parameterize it by
the mixture of only three components\footnote{We rely on the Occam's razor principle.}
\begin{equation}\label{eq:real-frequency-distribution}
\begin{split}
	f\left(\log\left(\nu_{i}\right)\mid  T_{\nu_{i}},\mu_{\nu},\boldsymbol{\theta}_{\nu},\tau_{\nu},\boldsymbol{\omega}_{\nu}\right) &  =    \delta_{T_{\nu_{i}}}^{1}\mathsf{N}\left(\mu_{\nu},\tau_{\nu}\right)+ \\ & \delta_{T_{\nu_{i}}}^{2}\mathsf{N}\left(\mu_{\nu}+\sqrt{\tau_{\nu}^{-1}}\theta_{\nu1},\tau_{\nu}\omega_{\nu1}^{-2}\right)+ \\	& \delta_{T_{\nu_{i}}}^{3}\mathsf{N}\left(\mu_{\nu}+\sqrt{\tau_{\nu}^{-1}}\theta_{\nu1}+\sqrt{\tau_{\nu}^{-1}}\omega_{\nu1}\theta_{\nu2},\tau_{\nu}\omega_{\nu1}^{-2}\omega_{\nu2}^{-2}\right) \;  .
\end{split}
\end{equation}
In Equation \ref{eq:real-frequency-distribution}, $\mu_{\nu}$ and
$\tau_{\nu}$ denote, respectively, the mean and the precision of the
first component of the
mixture. $\left(\theta_{\nu1},\theta_{\nu2}\right)$ and
$\left(\omega_{\nu1},\omega_{\nu2}\right)$ denote, respectively, the
perturbation parameters which affect the mean and the scale parameter
of a given component to obtain the mean and scale parameter of the
next component \cite{robert1999reparameterisation}. The Kronecker
deltas $\delta_{T_{\nu_{i}}}^{j}$ have the same role as in
Eq. \ref{eq:rec-frequency-density} but now the categorical variable
$T_{\nu_{i}}$ represents the class of the real frequency. For
$T_{\nu_{i}}$ we assign the distribution
\begin{equation}\label{eq:i-th-real-frequency-distribution}
	p\left(T_{\nu_{i}}\right)=\mathrm{\mathsf{Cat}}\left(3,w_{\nu1},w_{\nu2},w_{\nu3}\right) \;,
\end{equation}
where $w_{\nu j}$ are the mixing proportions of the mixture.

\paragraph{Input Amplitudes.}

 This distribution has been simulated based on the OGLE III
 catalogue of Classical Cepheids \cite{Soszynski2008}, as
 \begin{equation}\label{eq:Input-amplitude-distribution-theor}
 f\left(A\mid\log\left(\nu\right)\right)=\begin{cases}
 \mathsf{N}\left(-0.5\cdot\log\left(\nu\right)+0.2,0.15\right) & \log\left(\nu\right)<-1\\
 \mathsf{N}\left(0.7,0.15\right) & \log\left(\nu\right)>-1 \;  
 \end{cases}
 \end{equation}

 In our BGM we parameterize this variable as
 \begin{equation}\label{eq:Input-amplitude-distribution}
\begin{split}
f\left(A_{i}\mid\log\left(\nu_{i}\right),a_{A},b_{A},\mu_{A},\tau_{A}\right)&=  \boldsymbol{1}_{\left\{ \log\left(\nu_{i}\right)<-1\right\} }\mathsf{N}\left(a_{A}\cdot\log\left(\nu_{i}\right)+b_{A},\tau_{A}\right)\\
&+\boldsymbol{1}_{\left\{ \log\left(\nu_{i}\right)>-1\right\} }\mathsf{N}\left(\mu_{A},\tau_{A}\right)\;  ,
\end{split} 
\end{equation} 
 where $\boldsymbol{1}_{S}$ denotes the indicator function of a subset
 \emph{S}, $a_{A}$ and $b_{A}$ are, respectively, the slope and the
 intercept of the regression line of $A$ on $\log\left(\nu\right)$
 when $\log\left(\nu\right)<-1$, $\mu_{A}$ denotes the mean of the
 amplitude when $\log\left(\nu\right)>-1$, and $\tau_{A}$ denotes the
 precision, which we take equal in both cases.

\paragraph{Input Apparent G magnitudes.}

Based on Equations 12 and 13 of \cite{Sandage2004} and discarding the
distance $r$ to the sources, we parameterize this node as
\begin{equation}\label{eq:Input-apparent-G-magnitude-distribution}
\begin{split}
& f\left(m_{G,i}\mid\log\left(\nu_{i}\right),a_{G1},b_{G1},a_{G2},b_{G2},\tau_{G}\right)=\\
&\boldsymbol{1}_{\left\{ \log\left(\nu_{i}\right)<-1\right\} }\mathsf{N}\left(a_{G1}\cdot\log\left(\nu_{i}\right)+b_{G1},\tau_{G}\right)\\+&\boldsymbol{1}_{\left\{ \log\left(\nu_{i}\right)>-1\right\} }\mathsf{N}\left(a_{G2}\cdot\log\left(\nu_{i}\right)+b_{G2},\tau_{G}\right) \;  .
\end{split} 
\end{equation}

The conditional distribution of the first level of random parameters
given the parameters of the top level is then
\begin{equation}\label{eq:joint-PDF-1s-level-factor}
	\begin{split}
	p\left(\boldsymbol{\theta}_{1}\mid\boldsymbol{\theta}_{2}\right)&=\prod_{i=1}^{N}g_{1}\left(T_{\nu_{rec,i}}\mid\left\{
        \boldsymbol{\lambda}_{j}\right\} _{j=2}^{M}\right)\cdot
        g_{2}\left(A_{i}\mid\log\left(\nu_{i}\right),a_{A},b_{A},\mu_{A},\tau_{A}\right)\\\cdot
        &
        g_{3}\left(m_{G,i}\mid\log\left(\nu_{i}\right),\mathbf{a}_{G},\mathbf{b}_{G},\tau_{G}\right)\cdot
        g_{4}\left(\log\left(\nu_{i}\right)\mid
        T_{\nu_{i}},\lambda_{\nu},\boldsymbol{\theta}_{\nu},\tau_{\nu},\boldsymbol{\omega}_{\upsilon}\right)\\\cdot
        & g_{5}\left(T_{\nu_{i}}\mid\boldsymbol{w}_{\nu}\right)
	\end{split}
\end{equation}

\subsubsection{Top Level Random Parameters.}

These hyperparameters are
\begin{equation}\label{eq:model-top-level-parameters}
	\boldsymbol{\theta}_{2}=\left(a_{A},b_{A},\mu_{A},\tau_{A},\mathbf{a}_{G},\mathbf{b}_{G},\tau_{G},\mu_{\nu},\boldsymbol{\theta}_{\nu},\tau_{\nu},\boldsymbol{\omega}_{\nu},\boldsymbol{w}_{\nu},\Lambda\right)
        \; .
\end{equation}
	
$\boldsymbol{\theta}_{2}$ include the orphan nodes in the graph. We
only have a vague (or non informative) prior knowledge about their
distributions. The nodes denoted by $a$ and $b$ represent the slopes
and intercepts of the distributions of the real amplitude and apparent
G-magnitude given the frequency. The nodes denoted by $\tau$ and $\mu$
represent precisions and means. The nodes denoted by $\Lambda$
represent the coefficients of the logistic regression submodel of
Equation \ref{eq:softmax-recovery-probabilities}. The rest of nodes
are associated with the parameterization of the real frequency of
Equation \ref{eq:real-frequency-distribution}. For these latter
hyperparameters we take the non informative priors
\begin{align}\label{eq:prior-distributions-for-i-th-real-frequency}
  p\left(\boldsymbol{w}_{\nu}\right)&=\mathsf{Dir}\left(1,1,1\right) \\
  p\left(\mu_{\nu}\right)&=\mathsf{N}\left(0,0.001\right)\\
  p\left(\theta_{\nu j}\right)&=\mathsf{N}\left(0,0.01\right) \\
  p\left(\tau_{\nu}\right)&=\mathsf{Gamma}\left(0.001,0.001\right) \\
  p\left(\omega_{\nu j}\right)&=\mathsf{U}\left(0,1\right)
\end{align}

For the hyperparameters of the logistic regression submodel of
Equation \ref{eq:softmax-recovery-probabilities}
$\boldsymbol{\lambda}_{j}=\left(\lambda_{0j},\lambda_{1j}\right)$ 
  with $j\in\left\{ 2,...,M\right\} $, we assign the weakly
informative priors
$p\left(\lambda_{kj}\right)=\mathsf{t}\left(0,\frac{1}{2.5^{2}},7\right)$,
$k\in\left\{ 0,1\right\} $. This election provides a minimal prior
information to constrain the range of coefficients $\lambda_{kj}$ once
the covariate $\beta_{i}$ has been rescaled \cite{Gelman2008}. This
approximation is used to enhance the convergence rate of our model.

For the parameters $a_{A}$ , $b_{A}$ , $\lambda_{A}$ of the input
amplitude distribution of Equation
\ref{eq:Input-amplitude-distribution} and the parameters $a_{G1}$,
$b_{G1}$, $a_{G2}$ ,$b_{G2}$ of the input apparent G magnitude of
Equation \ref{eq:Input-apparent-G-magnitude-distribution} we take
$\mathsf{N}\left(0,0.001\right)$ non informative priors. And for the
precisions $\tau_{A}$ and $\tau_{G}$ we take
$\mathsf{Gamma}\left(0.001,0.001\right)$ priors. For all these priors the
full conditional distribution of the node is available in closed form.

The distribution (hyperprior) of the top level parameters is then
\begin{equation}\label{eq:joint-PDF-top-level-factor}
  \begin{split}
    p\left(\boldsymbol{\theta}_{2}\right)& =h_{1}\left(a_{A}\right)\cdot h_{2}\left(b_{A}\right)\cdot h_{3}\left(\mu_{A}\right)\cdot h_{4}\left(\tau_{A}\right)\cdot h_{5}\left(\mathbf{a}_{G}\right)\cdot h_{6}\left(\mathbf{b}_{G}\right) \cdot
    h_{7}\left(\tau_{G}\right)\\ &\cdot h_{8}\left(\boldsymbol{w}_{\nu}\right)\cdot h_{9}\left(\mu_{\nu}\right)\cdot h_{10}\left(\boldsymbol{\theta}_{\nu}\right)\cdot h_{11}\left(\tau_{\nu}\right)\cdot h_{12}\left(\boldsymbol{\omega}_{\nu}\right)\cdot h_{13}\left(\Lambda\right) \;  .
  \end{split}
\end{equation}

\subsubsection{Joint distribution of the Parameters and Data.}

From Equations \ref{eq:joint-PDF-likelihood-factor},
\ref{eq:joint-PDF-1s-level-factor} and
\ref{eq:joint-PDF-top-level-factor} we formulate the joint PDF
associated to the graphical mode by
\begin{equation}\label{eq:joint-PDF-model-factorization}
p\left(\boldsymbol{\theta},\mathcal{D}\right)=p\left(\mathcal{D}\mid\boldsymbol{\theta}\right)\cdot p\left(\boldsymbol{\theta}\right)=\\
p\left(\mathcal{D}\mid\boldsymbol{\theta}_{1}\right)\cdot p\left(\boldsymbol{\theta}_{1}\mid\boldsymbol{\theta}_{2}\right)\cdot p\left(\boldsymbol{\theta}_{2}\right) \;  . 
\end{equation}

\subsection{Computation}

The joint posterior distribution of the $22+5N$ parameters of the
  model described in Section \ref{sec:BGM} is given by
\begin{equation}\label{eq:joint-posterior-PDF-model-for-our-BGM}
\pi^{*}\left(\boldsymbol{\theta}\right)=\pi\left(\boldsymbol{\theta}\mid\mathcal{D}\right)\propto\mathcal{L}\left(\boldsymbol{\theta}_{1}\right)\cdot p\left(\boldsymbol{\theta}_{1}\mid\boldsymbol{\theta}_{2}\right)\cdot p\left(\boldsymbol{\theta}_{2}\right) \;  .
\end{equation}

Our goal is to infer the marginal \emph{a posteriori} distribution
$\pi^{*}\left(\boldsymbol{\theta}_{2}\right)$ of the top level
hyperparameters\footnote{In the case of the logarithm of the frequency
  distribution $\log\left(\nu\right)$ we are interested in the means
  and standard deviations of each Gaussian component, but obtaining
  these parameters from those in Equation
  \ref{eq:real-frequency-distribution} by deterministic relationships
  is straightforward.}.  The marginalization to obtain samples from
$\pi^{*}\left(\boldsymbol{\theta}_{2}\right)$ can be accomplished by a
general MCMC procedure in which, once a sample for the joint posterior
has been obtained, the procedure retains only the values of
$\boldsymbol{\theta}_{2}$ and discards the rest. The joint posterior
distribution of Equation \ref{eq:joint-posterior-PDF-model-for-our-BGM} can
be efficiently sampled by means of a Gibbs sampling scheme (see
Sec. 4.2 of \cite{Lunn2012}). To reduce our model to the programming
language level we have used the BUGS \cite{Lunn2009} probabilistic
language and the OpenBUGS software environment.

\section{{\bf Application to the {\it Gaia} Simulated Database of Classical Cepheids}}\label{sec:BGM-validation}

\begin{figure}
	\centering
	\includegraphics[scale=0.35]{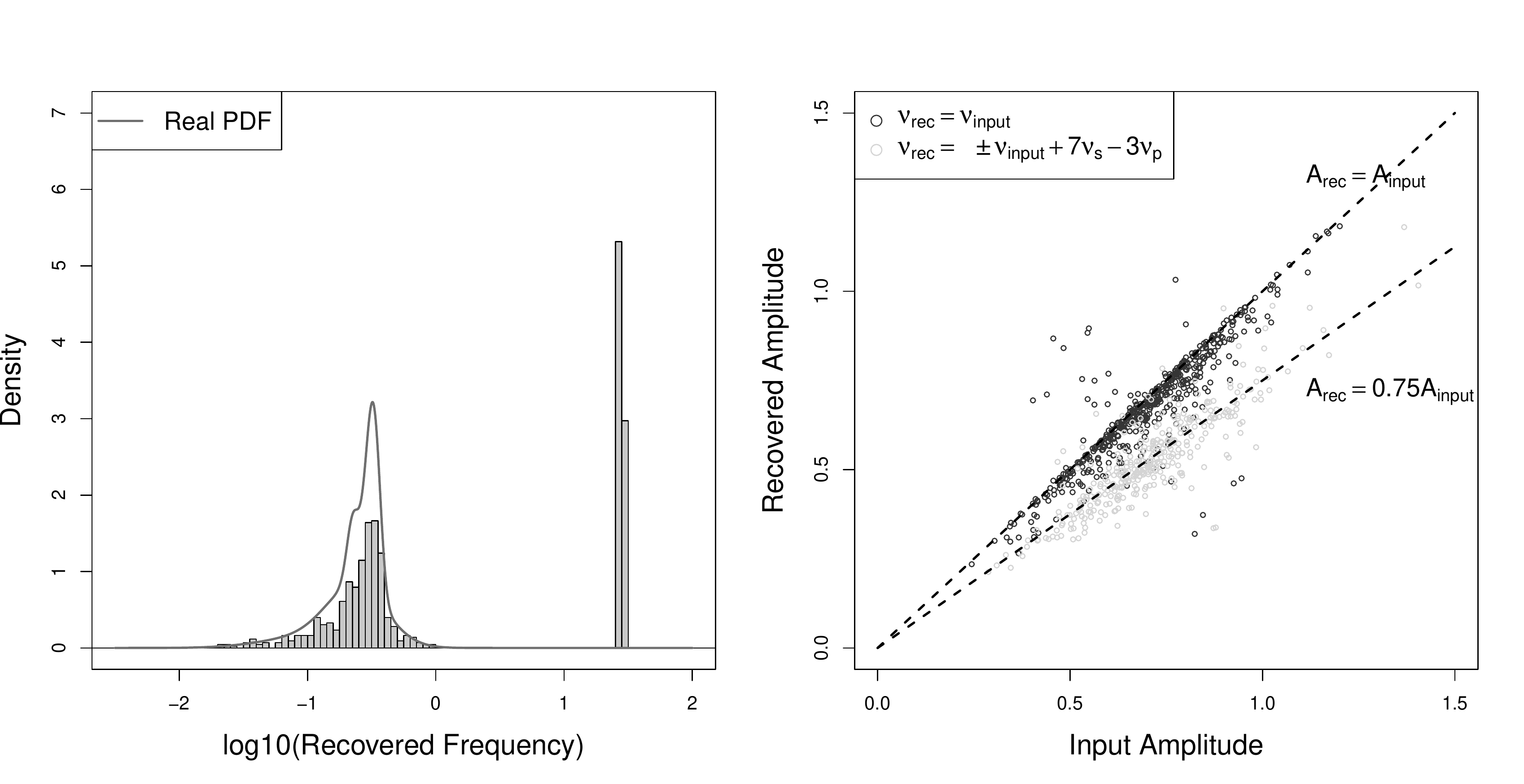}
	\caption{Biases in the frequencies (left) and amplitudes
          (right) present in the training set.}
	\label{fig:Biases-in-the-training-set}	
\end{figure}

In this Section we evaluate the effectiveness of our model to infer
the real distributions of hyperparameters in an extreme scenario
  of systematic biases in the recovered data. In order to do
  so, we have constructed a dataset {\footnotesize
  $\mathcal{T}=\left\{
  \left(A_{\mathrm{rec},i},\nu_{\mathrm{rec},i},m_{G,\mathrm{rec},i}\right)\right\}
  _{1}^{854}\varsubsetneq\mathcal{D}$ } composed of 500 randomly
  selected instances from the \emph{locus} {\footnotesize
  $\nu_{\mathrm{rec}}=\nu_{\mathrm{in}}$ }and all instances (354) from
the \emph{locus} {\footnotesize
  $\nu_{\mathrm{rec}}=\pm\nu_{\textrm{in}}+7\nu_{S}-3\nu_{p}$}.
Figure \ref{fig:Biases-in-the-training-set} shows the systematic
biases for the empirical frequency distribution (histogram) vs the
true probability density function (PDF) and for the empirical
conditional distributions of the recovered amplitude given the input
amplitude for the three  \emph{loci} (the identity locus and the
{\footnotesize
  $\nu_{\mathrm{rec}}=\pm\nu_{\textrm{in}}+7\nu_{S}-3\nu_{p}$ \emph{loci}}),
whose observed parameters are included in the training set.

We have trained the  model using the OpenBUGS MCMC engine.
We have divided the training in two stages and generated three Markov
chains (more properly realizations) in each, with a total of 30000
iterations.  We have used the first 20000 iterations as a
  \emph{burn-in} phase, and discarded them after using them for
  convergence assessment. Thereafter, we obtain 10000 samples
  from each chain in a second stage (30000 in total). We will assume
that these samples were drawn from the posterior distribution of the
parameters of interest.

\begin{table}[ht]
	\caption{Summary Statistics of Parameters of Interests. \label{tab:Statistics}}
	\centering
	\begin{tabular}{cccccc}
		\hline 
		{$\theta$} & $\overline{\mathrm{ACR}}$ &  GRB & $\overline{\mathrm{\theta}}$ &  2.5\%-97.5\% Perc. & Real value \tabularnewline 
		\hline 
		$w_{\nu2}$ & 0.39 & 1.09 & 0.03  & 0.01,0.05 & -\tabularnewline 
		$w_{\nu1}$ & 0.18 & 1.03 & 0.41 & 0.32,0.50 & -\tabularnewline
		$w_{\nu3}$ & 0.16 & 1.01 & 0.57  & 0.47,0.66 & -\tabularnewline
		\hline 
		$\mu_{\nu2}$ &  0.60 &  1.16 & -1.50 & -1.61,-1.37 & - \tabularnewline
		$\mu_{\nu1}$ & 0.19 & 1.01 & -0.66  & -0.71,-0.61 & - \tabularnewline
		$\mu_{\nu3}$ & 0.05 & 1.01 & -0.53 & -0.54,-0.51 & -\tabularnewline
		\hline 
		$\sigma_{\nu2}$ & 0.83 & 1.25 & 0.14 & 0.10,0.20 & - \tabularnewline
		$\sigma_{\nu1}$ & 0.25 & 1.01 & 0.28 & 0.25,0.33 & - \tabularnewline
		$\sigma_{\nu3}$ & 0.09 & 1.02 &  0.09  & 0.08,0.10 & - \tabularnewline
		\hline 
		$a_{A}$ & 0.04 &  1.00 & -0.43 & -0.67,-0.21  &  -0.5\tabularnewline
		$b_{A}$ & 0.03 & 1.00 &  0.28 &  -0.02,0.57  &  0.2\tabularnewline
		$\mu_{A}$ & 0.00  & 1.00 &  0.62 &  0.58,0.66  &  0.7\tabularnewline
		$\sigma_{A}$ & 0.00 & 1.00 & 0.15 & 0.14,0.16 & 0.15\tabularnewline
		\hline 
		$a_{G1}$ & 0.25 & 1.04 & 2.55 & 2.22 ,2.91 & -\tabularnewline
		$b_{G1}$ & 0.23 & 1.03 & 16.76 & 16.38 ,17.17 & -\tabularnewline
		$a_{G2}$ & 0.01 & 1.00 & 3.01  & 2.96 ,3.06 & -\tabularnewline
		$b_{G2}$ & 0.01 & 1.00 &  17.16 & 17.13 ,17.19 & -\tabularnewline
		$\sigma_{G}$ &  0.00  & 1.00 & 0.10  &  0.09,0.11 & -\tabularnewline
		\hline
		$\lambda_{02}$ & 0.00 & 1.00 &  -1.132  & -1.314 ,-0.955 & -\tabularnewline
		$\lambda_{03}$ &  0.00 & 1.00 & -0.981  & -1.156 ,-0.816 & -\tabularnewline
		$\lambda_{\beta2}$ & 0.00 & 1.00 & -0.766 & -1.140 ,-0.395 & -\tabularnewline
		$\lambda_{\beta3}$ &  0.00 &  1.00 &  -0.743  & -1.091 ,-0.385 & - \tabularnewline
		\hline	
	\end{tabular}
\end{table}

\subsection{Convergence analysis}
To evaluate the convergence within and between the three chains we
have selected the first 20000 iterations of the algorithm and computed
the mean autocorrelation (ACR) (after 200 lags) and the upper bound of
a credible interval (at 95\%) for the corrected GR statistic
\cite{Brooks1998}. The results of the analysis are  summarized in
the second and third columns of Table \ref{tab:Statistics}. Since the
ACR function should decrease to zero as the lag increases and the
upper bound for the corrected scale reduction factor (CSRF) should
approach unity if the chain is reaching its stationary distribution,
we conclude that the worst scenario (high autocorrelation) is 
  encountered in the chains of the parameters specifying the second
Gaussian component of $\log\left(\nu\right)$, namely the mixing
proportion $w_{\nu2}$, the mean $\mu_{\nu2}$ and the standard
deviation $\sigma_{\nu2}$. In particular, chains for $\sigma_{\nu2}$
show the worst behaviour with a mean ACR after 200 lags of about 0.8
and a CSRF upper bound of 1.25. In contrast, the best scenario is 
  found in the chains of the parameters of the conditional
distributions of apparent G-magnitude and amplitude (given the
frequency) when $\log\left(\nu\right)>-1$, and by chains of logit
coefficients. For the slope $a_{G2}$, the intercept $b_{G2}$, the mean
$\mu_{A}$ and the logit coefficients $\lambda_{\beta
  j}$,$\lambda_{0j}$, $j\in\left\{ 1,2\right\}$ the mean ACR is nearly
zero after lags greater than 50 and the CSRF bound is close to unity.

\subsection{Posterior Distributions and Comparison with Real Parameters}
In this Section we evaluate the ability of our model to retrieve the
real distributions of the frequency, amplitude and apparent
G-magnitude of the simulated Cepheids sample from the recovered values
in the training set {\footnotesize $\mathcal{T}$}. We first
  compute summary statistics (means and 2.5\%-97.5\% percentiles) for
the samples of the posterior distributions of the hyperparameters
inferred by the model. Then, we have compared the posterior means with
the parameters of the real theoretical distributions used to generate
the simulated sample. Finally, we have constructed theoretical
distributions using the posterior means and compared them with the
true theoretical distributions and the empirical distribution in the
set {\footnotesize $\mathcal{I}=\left\{
  \left(A_{\mathrm{in},i},\nu_{\mathrm{in},i},m_{G,\mathrm{in},i}\right)\right\}
  _{1}^{854}$} .

\begin{figure}
	\centering
	\includegraphics[scale=0.34]{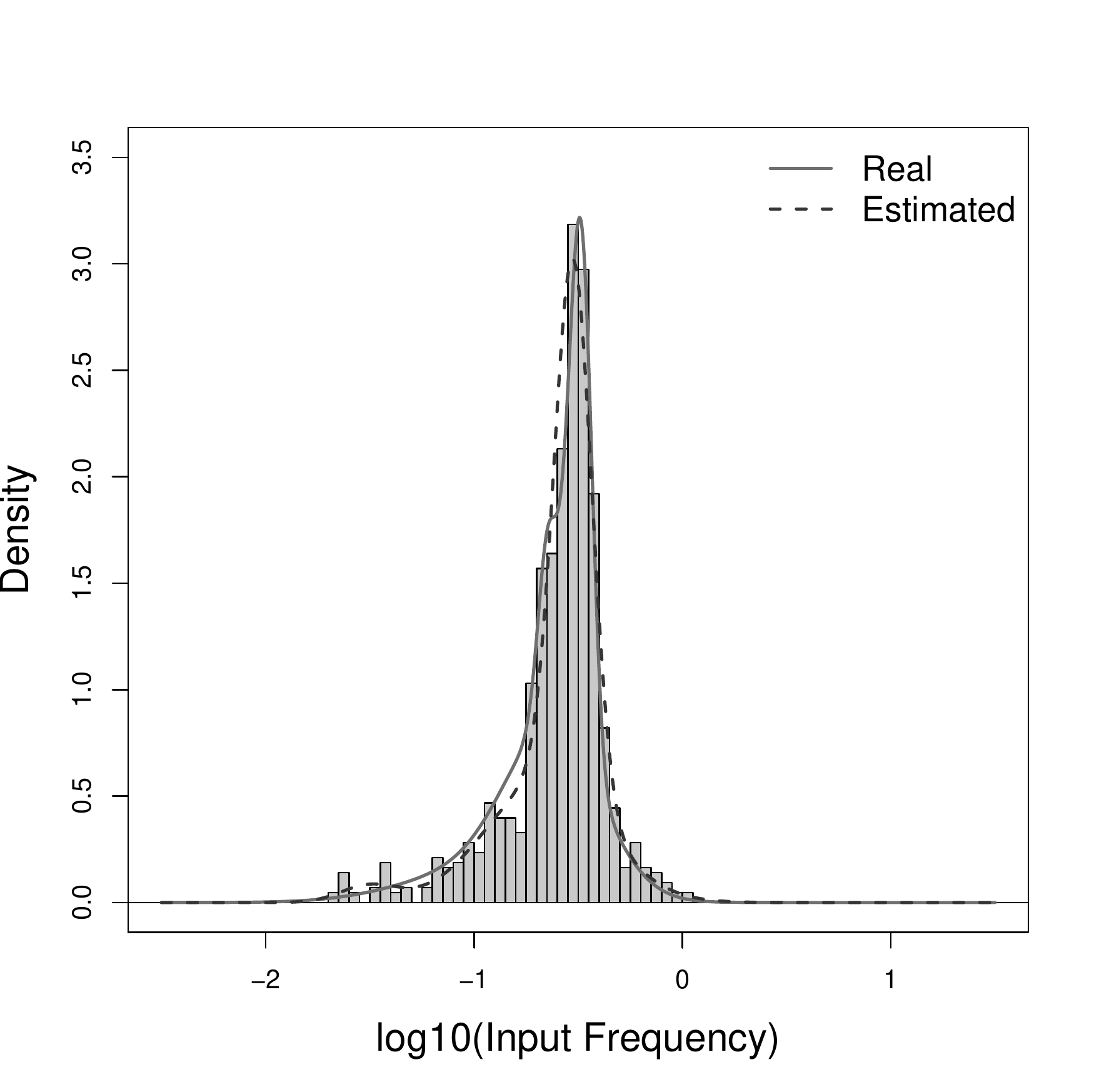}
	\includegraphics[scale=0.34]{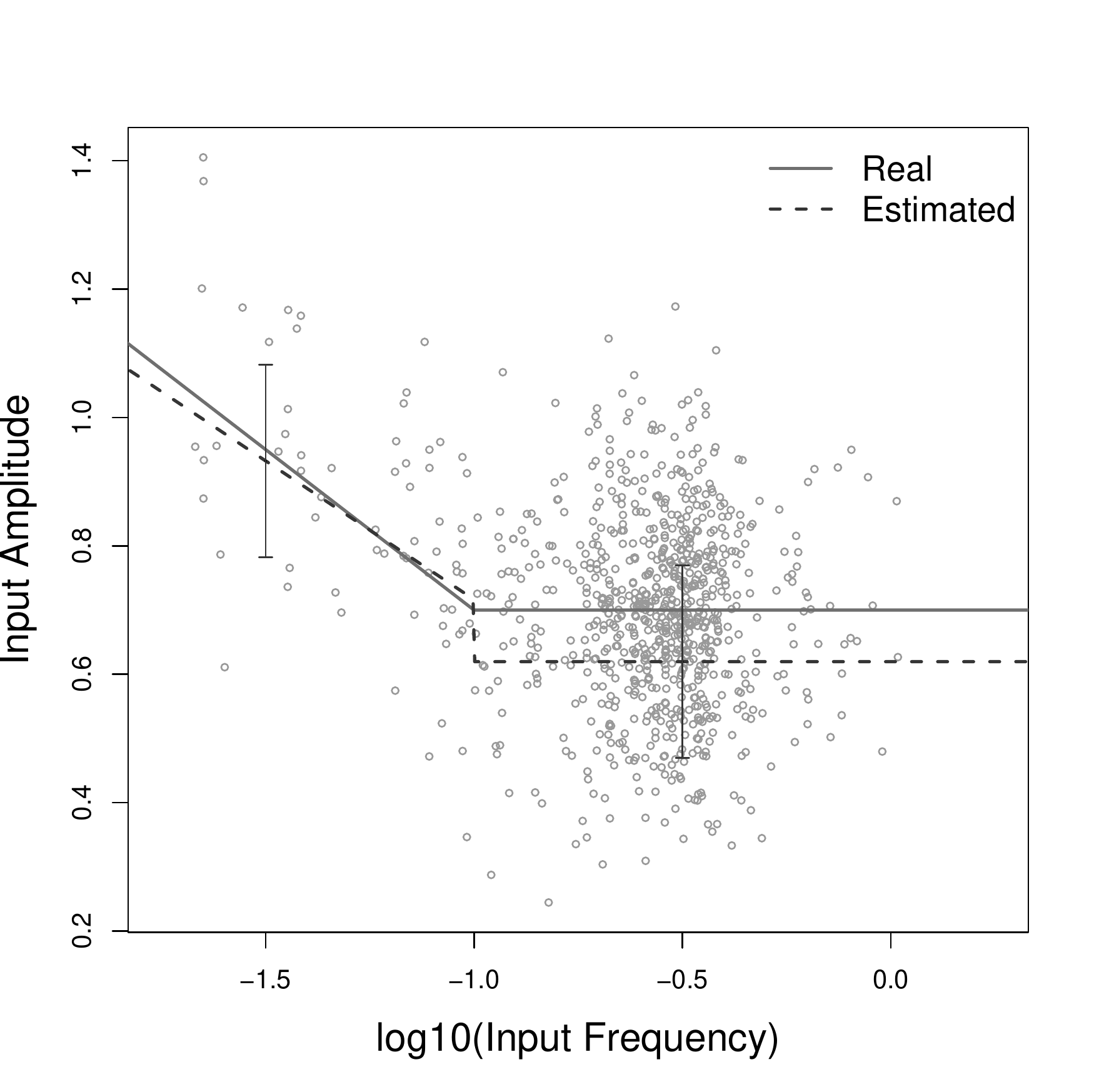}
	\caption{Posterior versus Real Distributions.}
	\label{fig:posterior-vs-real-distributions}	
\end{figure}

The results of our analysis are shown in Table \ref{tab:Statistics}
and Figure \ref{fig:posterior-vs-real-distributions}. We do not
include in the Table the parameters used to generate the real
frequency $\log\left(\nu\right)$, because it is difficult to make a
correspondence with the inferred parameters due to the different
number of Gaussian components. But if we observe the comparison graph
to the left of Figure \ref{fig:posterior-vs-real-distributions}, we
conclude that the fitting of $\log\left(\nu\right)$ with three
components (dotted line), reconstructs the real PDF (solid line)
successfully.

For the parameters of the conditional distribution
$A_{\textrm{in}}\mid\log\left(\nu_{\textrm{in}}\right)$ we fitted the
piecewise linear model of Equation
\ref{eq:Input-amplitude-distribution}. The middle rows of Table
\ref{tab:Statistics} and the graph at the right of Figure
\ref{fig:posterior-vs-real-distributions} show that the system
underestimates the true value of the mean $\mu_{A}$ when
$\log\left(\nu_{\textrm{in}}\right)>-1$.

\section{Summary and Conclusions}\label{sec:conclusions}

We have presented a two-level BGM to infer the real distributions of
amplitude, frequency and apparent G-magnitude of the Large Magellanic
Cloud population of Classical Cepheids from the values recovered by
the {\it Gaia} DPAC pipeline. We have modelled the real frequency by a
mixture of three Gaussian distributions and used piecewise linear
models (with a fixed knot value depending on the frequency) to model
the dependency of the true amplitude and G-magnitude on the true
frequency. We have tackled the problem of aliasing in the DPAC
frequency recovery module which arises as a result of the Gaia
scanning law. We have modelled the recovery probabilities in various
\emph{loci} of aliased frequencies using a logistic regression submodel based
on the ecliptic latitude predictor. We have modelled the recovered
frequencies and amplitudes as generated from mixtures of distributions
where the mixing proportions are the recovery probabilities. Although
our model has not yet solved completely the aliasing problem (we have
only used some predefined configurations of aliased data, and we have
restricted the application to a very narrow range of ecliptic
latitudes in which the relationship between the recovery probability
of aliased frequencies and the ecliptic latitude is monotone) it
represents a major step forward. The next step will necessarily
consist in extending the analysis to the full celestial sphere by
clustering the full variety of time samplings (and corresponding
window functions) into discrete bands of ecliptic longitudes and
latitudes.

\bibliographystyle{splncs03}
\bibliography{D-S-Iwinac17}

\end{document}